\documentclass[
reprint,
superscriptaddress,
 amsmath,amssymb,
 aps,
]{revtex4-2}

\usepackage{graphicx}% Include figure files
\usepackage{dcolumn}% Align table columns on decimal point
\usepackage{bm}% bold math
\usepackage{xcolor}
\usepackage{placeins}
\begin{document}

%\preprint{APS/123-QED}

\title{Field-dependent Magnons in a Honeycomb Antiferromagnet CoTiO$_3$}

\author{Bo Yuan}
\thanks{Current affiliation: Department of Physics and Astronomy, McMaster University, Hamilton, Ontario L8S 4M1, Canada}
\address{Physics Department, University of Toronto, 60 St. George Street, Toronto, Ontario M5S 1A7, Canada}
\author{Ezekiel Horsley}
\address{Physics Department, University of Toronto, 60 St. George Street, Toronto, Ontario M5S 1A7, Canada}
\author{M.B. Stone}
\address{Neutron Scattering Division, Oak Ridge National Laboratory, Oak Ridge, Tennessee 37831, USA}
\author{Nicholas P. Butch}
\address{NIST Center for Neutron Research, National Institute of Standards and Technology Gaithersburg, Maryland, 20899, USA}
\author{Guangyong Xu}
\address{NIST Center for Neutron Research, National Institute of Standards and Technology Gaithersburg, Maryland, 20899, USA}
\author{Guo-Jiun Shu}
\address{Department of Materials and Mineral Resources Engineering, National Taipei University of Technology, Taipei 10608, Taiwan}
\address{Institute of Mineral Resources Engineering, National Taipei University of Technology, Taipei 10608, Taiwan}
\address{Taiwan Consortium of Emergent Crystalline Materials, Ministry of Science and Technology, Taipei 10622, Taiwan}
\author{J.P. Clancy}
\address{Department of Physics and Astronomy, McMaster University, Hamilton, Ontario L8S 4M1, Canada}
\author{Young-June Kim}
\address{Physics Department, University of Toronto, 60 St. George Street, Toronto, Ontario M5S 1A7, Canada}

\date{\today}
\begin{abstract}
We report field-dependent high-resolution inelastic neutron scattering (INS) measurements on the honeycomb lattice magnet, CoTiO$_3$, to study the evolution of its magnon excitations across a spin reorientation transition driven by an in-plane magnetic field. By carrying out elastic neutron scattering in a magnetic field, we show that the sample transitions from a collinear antiferromagnetic state with multiple magnetic domains at a low field to a mono-domain state with a canted magnetic structure at a high field. Concurrent with this transition, we observed significant changes in both the energy and the width of the zone center magnon peak. The observed width change is argued to be consistent with an unusual zero-field state with extended domain walls. On the other hand, the magnon spectra near the $\mathbf{K}$ point of the Brillouin zone boundary are found to be largely insensitive to the changes in the ordered moment directions and the domain configuration. We argue that this observation is difficult to explain within the framework of the bond-dependent model proposed in a recent INS study [Elliot \textit{et\,al}, Nat. Commun., \textbf{12}, 3936 (2021)]. Our study therefore calls for alternative explanations for the observed $\mathbf{K}$-point gap in CoTiO$_3$.  
\end{abstract}

\maketitle
\section{Introduction}\label{Intro}
Searching for magnetic materials with dominant Kitaev interactions\citep{KITAEV20062}, or more generally bond-dependent interactions, have become an important goal of today's research in quantum magnetism due to their ability to support exotic ground states such as the quantum spin liquid state. After more than a decade of intensive experimental and theoretical research, two properties, namely, strong spin-orbit coupling and an edge-sharing bond geometry have been identified as the key ingredients for hosting such non-trivial exchange interactions\citep{Jackeli2009}. With these general considerations, a few promising examples have emerged including $\alpha-$RuCl$_3$, iridates with honeycomb and hyper-honeycomb structure, and more recently a number of honeycomb cobaltates (Given the large number of works in this still rapidly developing field, instead of giving references on specific materials, here we only cite a number of review articles. See Ref.~\citep{Motome_2020,TREBST20221,Winter_2017,Hermanns2018,Takagi2019,Liu2021review,Subin2022}) A wealth of experimental results ranging from magnetic order (the so called zigzag order in a honeycomb lattice has been widely associated with a dominant Kitaev term\citep{Chaloupka2013}) to some unconventional bulk~\citep{Sears2020,Kelley2018,Lin2021Na2Co2TeO6}) and thermal transport responses~\citep{Kasahara2018,Yokoi2021,Czajka2023,Czajka2021}) have been cited as indirect evidence for the existence of such interactions and possible exotic ground states in these materials. However, a quantitative understanding of the spin Hamiltonian has not been achieved in most of them. This is the case even for the most well-known Kitaev candidate RuCl$_3$ where the magnitude and the form of the dominant bond-dependent interactions are still under scrutiny (E.g. See Table I of Ref.~\citep{MaksimovPRR2020} for a summary of all bond-dependent interactions proposed for RuCl$_3$)

The main difficulty is a lack of careful measurement of their magnon excitations (either in the zero-field ordered phase, or a field polarized phase when the zero-field excitation spectrum is strongly damped such as $\alpha-$RuCl$_3$\citep{RuCl3neutron}) that can be compared with linear spin wave calculations - the gold standard for parametrization of the spin Hamiltonian in a material. Compared to $\alpha$-RuCl$_3$ and the honeycomb iridates both with a small moment size, and in the case of iridates, a large neutron absorption cross-section, honeycomb cobaltates are clearly better suited for such purposes because of the strong magnetic scattering by the large Co$^{2+}$ moments. Inelastic neutron scattering (INS) measurements carried out on powder samples of honeycomb cobaltate all show well-defined spin waves with a finite anisotropy gap, consistent with the existence of some type of exchange anisotropy between the pseudospins, although the extracted exchange parameters can vary significantly when trying to fit the powder averaged spectrum\citep{Nair2018,Lin2021Na2Co2TeO6,Songvilay2020,Sanders2022,Kim_2022,Samarakoon2021}. On the other hand, more detailed INS measurements of magnon excitations using single crystal samples became available only recently for two of the most prominent candidates for dominant Kitaev interactions, Na$_2$Co$_2$TeO$_6$\citep{Yao2022} and BaCo$_2$(AsO$_4$)$_2$\citep{Halloran2023}. Surprisingly, the single crystal magnon spectra for both of these compounds show large deviations from the predictions of a simple model with dominant Kitaev interactions, at odds with some early powder INS work\citep{Songvilay2020,Kim_2022,Samarakoon2021,Sanders2022}. These recent results raise important questions about the relevance of bond-dependent interactions \citep{Liutheory2020,Liutheory2018} in real materials, and call for careful modelling of their single crystal magnon spectra.

\begin{figure*}[hbt]
\includegraphics[width=1\textwidth]{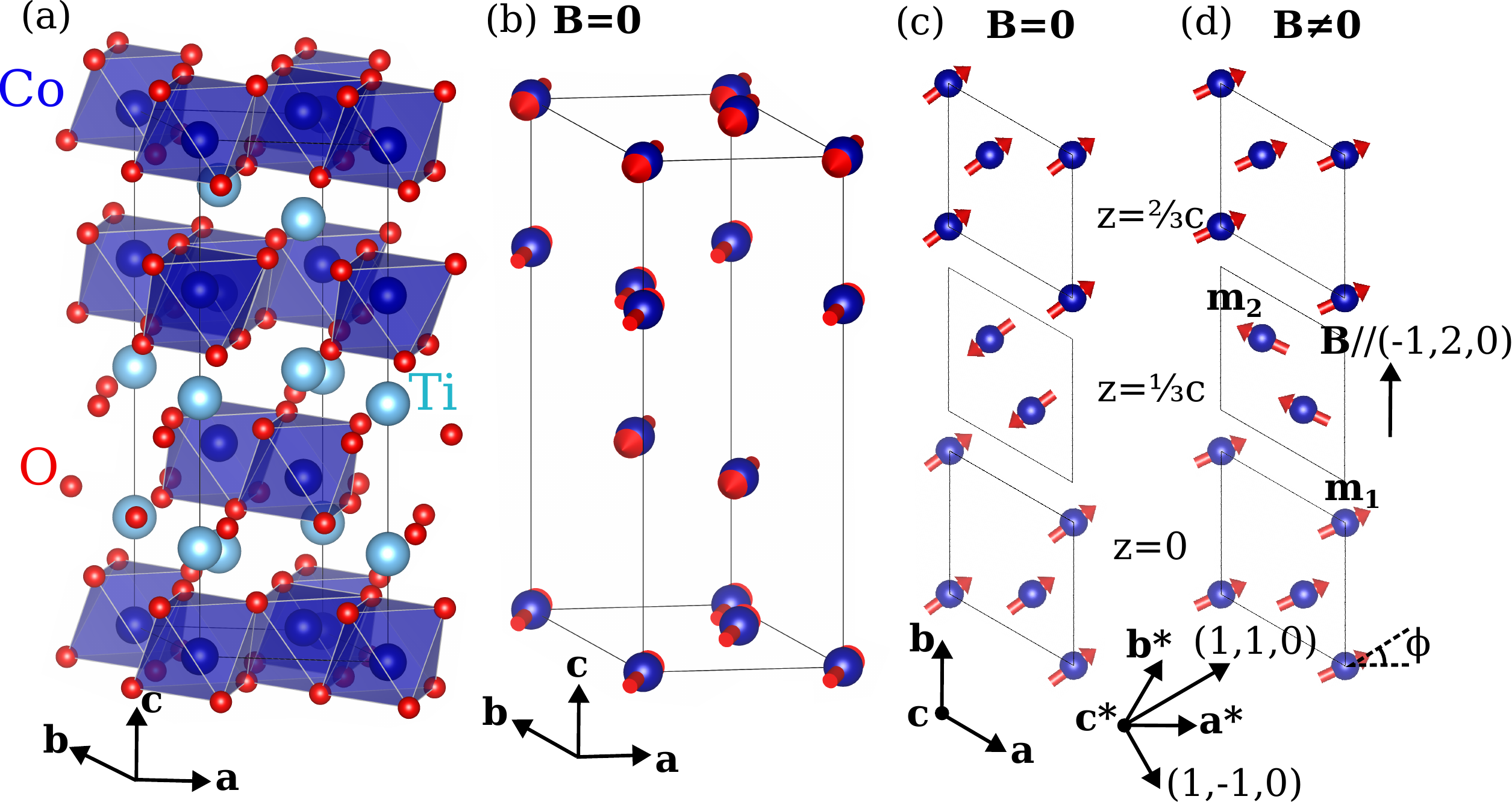}
\caption{\label{structure} (a,b) 3-dimensional crystal and magnetic structure of CoTiO$_3$. (c) ordered moment configuration for three consecutive Co honeycomb layers within each structural unit cell separated by $\frac{1}{3}c$ at zero field. (d) is the same as (c) but in the high-field mono-domain state for a field applied along the crystallographic $\mathbf{b}$ axis or equivalent (-1,2,0) in reciprocal space. At zero field, the ordered moments on neighbouring honeycomb layers denoted by $\mathbf{m}_1$ and $\mathbf{m}_2$, respectively, are collinear and are antiferromagnetically coupled. Without the knowledge of the actual ordered moment directions at zero field, the ordered moments are shown to point along an arbitrary in-plane direction in (b) and (c).  At high field, the two antiferromagnetic sublattices cant away from the collinear configuration towards the direction of the applied field. The size of the canting is denoted by $\phi$ in (d).  The trigonal unit cell used throughout the paper has been indicated by the black solid lines. The relevant in-plane real-space and reciprocal space coordinates are indicated by black solid arrows in (c) and (d), respectively.}
\end{figure*}

To address these questions, we revisit the magnon dispersions in a well-known honeycomb lattice with edge-sharing bond geometry, CoTiO$_3$. Unlike most Kitaev candidates with a zigzag order, the ground state of CoTiO$_3$ has a simple A-type antiferromagnetic order with ferromagnetically ordered honeycomb planes antiferromagnetically coupled in the out-of-plane direction (See Fig.~\ref{structure} for the crystal and magnetic structure of CoTiO$_3$). Simplicity of the magnetic order in CoTiO$_3$ allowed detailed modelling of the magnon spectra of single crystal samples~\citep{BoYuan2020PRX, Elliot2021}. Our first INS measurement\citep{BoYuan2020PRX} found a magnon spectrum consistent with a dominant XXZ interaction between the pseudospins, while subsequent high resolution INS\citep{Elliot2021} and optical measurements\citep{li2022ringexchange} discovered subtle discrepancies that require exchange anisotropies beyond the XXZ model. First, like the other honeycomb cobaltates, a gap in the Goldstone mode indicative of an in-plane U(1) symmetry breaking was discovered at the magnetic zone center, requiring either a quantum order by disorder mechanism with bond-dependent interactions \citep{Elliot2021} or a multi-spin ring exchange interaction \citep{li2022ringexchange}. In addition, a gap at the $\mathbf{K}$ point of the Brillouin zone (also referred to as a Dirac gap) was discovered between the optical and acoustic magnon branches, which was modelled using a linear spin wave theory with the bond-dependent interactions \citep{Elliot2021}. 

However, there are two difficulties in the modelling of the zero-field spectra that render the determination of the exchange parameters in CoTiO$_3$ an under-constrained problem. First, the directions of ordered moments necessary for any linear spin wave calculations are not known in CoTiO$_3$, which had to be assumed in Ref~\citep{Elliot2021}. Second, there exist both structural and magnetic domains at zero field in CoTiO$_3$. The structural domains are related by a two-fold rotation around (1,1,0), while the configuration of magnetic domains at zero field is unknown in CoTiO$_3$. Since different domains give rise to distinct excitation spectra, knowledge of the actual domain configuration is important for the correct simulation of the zero-field spectra which is an average over all domains.
 
In this paper, we address these difficulties by carrying out field-dependent high resolution neutron scattering measurements with an in-plane magnetic field. The in-plane field drives a spin reorientation transition from a zero-field multi-domain state to a high-field mono-domain state with a canted magnetic structure in CoTiO$_3$, which we characterize using elastic neutron scattering. The changes in magnon spectra near both the magnetic zone center (ZC) and the \textbf{K} point of the Brillouin zone boundary were studied by high resolution INS measurements with a magnetic field. In addition to a change in the energy of the ZC magnon, we observe a strong suppression of its width across the spin reorientation transition, which suggests a complex zero-field state with extended domain walls. On the other hand, we find very similar magnon spectra near the $\mathbf{K}$ point in both the high-field mono-domain state and the zero-field multi-domain state, which we argue to be inconsistent with the behaviour expected for general bond-dependent interactions\citep{Elliot2021}. We close by discussing alternative explanations for the observed gap and possible experimental tests for them.
  
\section{Experimental Details}
The same CoTiO$_3$ crystal grown by the floating zone method from Ref~\citep{BoYuan2020PRX, BoYuan2020PRB} was used in the measurements reported in this paper. Inelastic neutron scattering (INS) measurements of the low energy zone center magnons were carried out at the Disk  
Chopper Spectrometer (DCS) and the Spin Polarized Inelastic Neutron Spectrometer (SPINS) at the NIST Center for Neutron Research (NCNR). In the DCS experiment, the crystal was aligned in the (H,H,L) scattering plane (Throughout the paper, we use a hexagonal unit cell with $a=b=5.0662\AA$ and $c=13.918\AA$). Two incident energies, $\mathrm{E_i}=4.04~\mathrm{meV}$ and $\mathrm{E_i}=5.66~\mathrm{meV}$, were used, which offered an energy resolution of $\mathrm{\Delta\mathrm{E}}\approx 0.13 \mathrm{meV}$ and $\mathrm{\Delta\mathrm{E}}\approx 0.23 \mathrm{meV}$, respectively, at the elastic line. In the SPINS experiment, the crystal was aligned in the (H,0,L) plane and a fixed final energy, $\mathrm{E_f}=5~\mathrm{meV}$, was used. An energy resolution of $\sim$0.2~meV was achieved with a vertically focussing pyrolitic graphite (PG) monochromator, a flat PG analyzer, a Be filter and a collimation setting of guide-open-80'-open. The same 10~T vertical field magnet was used in the DCS and SPINS experiments to apply a field along the (1,-1,0) and (-1,2,0) directions, respectively. The high resolution measurements of the high energy magnon near the $\mathbf{K}$ point were carried out at the fine-resolution Fermi chopper spectrometer, SEQUOIA, at the Spallation Neutron Source (SNS) at the Oak Ridge National Laboratory. The crystal was aligned in the (H,H,L) plane. A T0 chopper and a high-resolution Fermi chopper, FC2, rotating at frequencies of 30~Hz and 180~Hz, respectively, were used to select an incident energy of 12~meV with a resolution of $\sim$0.27~meV at the elastic line. An 8~T vertical field magnet was used to apply a field along (1,-1,0). Due to the lack of radial collimation at SEQUOIA, the magnet gave a large background that was removed by taking empty magnet measurements without the sample. Alignment of the crystal before all experiments was carried out at the McMaster Alignment Diffractometer (MAD) at the McMaster Nuclear Reactor at McMaster University. Magnetization measurements were carried out using a Quantum Design SQUID Magnetic Property Measurement System. Linear spin wave modelling of the INS data was performed using the SpinW package\citep{SpinW}. 

\section{Results}
\subsection{Spin Reorientation}\label{spinflop}
\begin{figure}[hbt]
\includegraphics[width=0.5\textwidth, trim=0cm 0cm 0cm 0cm,clip]{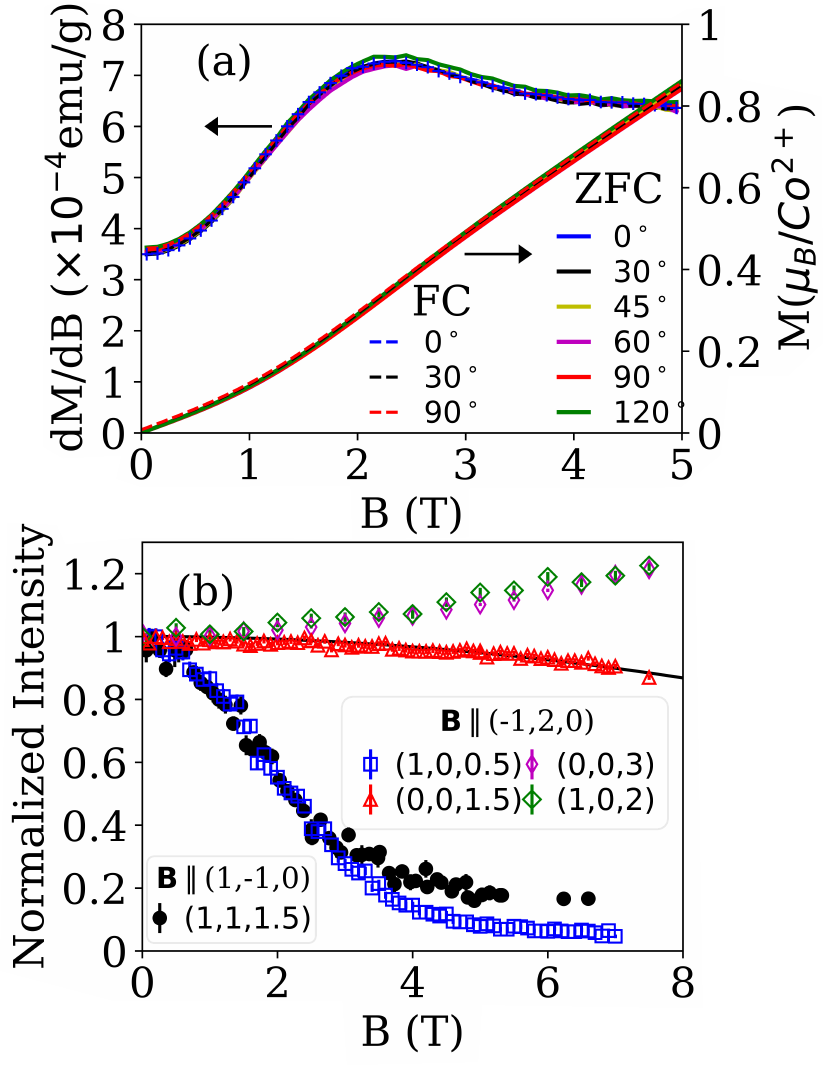}
\caption{\label{elastic} (a) Magnetization, $M$, and magnetic susceptibility, $\frac{dM}{dB}$, as a function of field strength for different field directions. The magnetic field was applied in the $\mathbf{ab}$-plane at different angles with respect to the crystallographic $\mathbf{a}$-axis. For each measurement, the sample was warmed to room temperature, rotated to the correct orientation with respect to the field, and cooled to 2~K either in zero field (ZFC) or in a 5~T field (FC). The M vs B measurement was then carried out from 0~T to 5~T. In the FC case, the field was first ramped down to 0~T before the start of the measurement. (b) Field dependence of magnetic and nuclear Bragg peak intensities for different field directions. Data shown in solid (open) symbols were obtained at DCS (SPINS) with field applied along the (1,-1,0) [(-1,2,0)] direction, respectively. For comparison, intensities of different Bragg peaks were normalized to their zero-field values. The solid line is a fit of the field dependence of the (0,0,1.5) peak is to a quadratic function as described in the text.}
\end{figure}

The magnetization as a function of magnetic field is plotted in Fig.~\ref{elastic}~(a) for several field directions. (Note that we use B to denote magnetic field and H is used to refer to the reciprocal lattice index along \textbf{a*}.) The magnetic susceptiblity $\frac{dM}{dB}$ is also plotted in the same figure. The spin reorientation in CoTiO$_3$ induced by an in-plane magnetic field was first observed through bulk magnetization measurements \citep{CoTiO3scsynthesis}. As shown in Fig.~\ref{elastic}~(a), the transition is gradual and manifests as a gradual increase in $\frac{dM}{dB}$ up to $\sim$2~T above which $\frac{dM}{dB}$ approximately doubles compared to that at zero-field.  The observed change is consistent with a transition from a zero-field multi-domain state favoured by the intrinsic spin anisotropy to a high field mono-domain state where the ordered moments are aligned approximately perpendicular to the applied field to maximize the Zeeman energy gained by canting of the antiferromagnetic sub-lattices [See Fig.~\ref{structure}~(d)]. 

Since only the spin component perpendicular to momentum transfer, $\mathbf{Q}$, is detected in a neutron diffraction experiment, we directly observed the change in ordered moment direction by tracking the field dependence of the intensity of a magnetic Bragg peak with a large in-plane $\mathbf{Q}$ component. In our experiments, a magnetic field was applied along a direction in the $\mathbf{ab}$ plane with a vertical field magnet by aligning the crystal in a scattering plane containing (0,0,1). For example, by aligning the crystal in the (H,0,L)-plane, the field is applied perpendicular to (1,0,0) along the crystallographic $\mathbf{b}$-axis [This is equivalent to (-1,2,0) in reciprocal space. In this section, we will use the real-space and reciprocal space notations interchangeably- See Fig.~\ref{structure} for their relations.]. Under this field, the ordered moments are mostly along the (1,0,0) direction in the high-field state [Fig.~\ref{structure}~(d)]. This leads to suppression of the (1,0,L)-type peaks with a small L, since $\mathbf{Q}$ is almost parallel to the ordered moments. This is consistent with the measured field dependence of the (1,0,0.5) magnetic Bragg peak (Fig.~\ref{elastic}~(b)) where an intensity drop mirrors the behaviour of $\frac{dM}{dB}$. On the other hand, the intensity of the (0,0,1.5) Bragg peak has a very weak field dependence, consistent with the fact that the spins only rotate in the $\mathbf{ab}$ plane, and are therefore always perpendicular to $\mathbf{Q}=(0,0,1.5)$. The (0,0,1.5) Bragg peak intensity decreases quadratically with field, as shown by a fit of its field dependence to $I_{(0,0,1.5)}=I_0\left[1-(\frac{B}{B_\mathrm{sat}})^2\right]$ (black solid line in Fig.~\ref{elastic}~(b)), where $B_\mathrm{sat}$ stands for the saturation field. The field dependence could be either due to a reduction in the magnitude of the staggered magnetization, $\mathbf{n}~=~\mathbf{m}_1-\mathbf{m}_2$, or a sample movement induced by a large applied field. $\mathbf{m}_1$ and $\mathbf{m}_2$ are the sub-lattice magnetizations on two adjacent honeycomb layers as shown in Fig.~\ref{structure}~(d). The sample movement is ruled out by the field dependence of the two nuclear Bragg peak intensities, (1,0,2) and (0,0,3). Unlike (0,0,1.5), their intensities \textit{increase} quadratically with the field. These observations can be explained by a canting of $\mathbf{m}_1$ and $\mathbf{m}_2$ away from their zero-field collinear configuration towards the applied field [see Fig.~\ref{structure}~(d)], leading to a decrease in the size of $\mathbf{n}=\mathbf{m}_1-\mathbf{m}_2$ but an increase in the net  magnetization or $\mathbf{m}=\mathbf{m}_1+\mathbf{m}_2$, the latter of which contributes to the elastic intensity at nuclear Bragg peak positions. 

To estimate the spin canting at high field, we note that $I_{(0,0,1.5)}(\mathrm{B})~\propto~m_0 (\mathrm{B})^2~\cos^2 [\phi(\mathrm{B})]$, where $\phi$ is the canting of sub-lattice magnetizations from $\mathbf{n}$, and $|\mathbf{m}_1|=|\mathbf{m}_2|=m_0$ is proportional to the size of the Co$^{2+}$ ordered moment. The field dependence of $m_0 (\mathrm{B})$ can be determined from the high field magnetization measurement above the saturation field when all moments have been polarized by the field \citep{Hoffmann2021highfield}. In Ref.~\citep{Hoffmann2021highfield}, $m_0$ increases linearly with field as $\frac{m_0(\mathrm{B})-m_0(0)}{m_0(0)}=0.0044 \mathrm{B}$ (B is in Tesla), which can be attributed to a field-induced admixture of the excited doublet ($J_z^{\mathrm{tot}}=\pm\frac{3}{2}$) into the ground doublet ($J_z^{\mathrm{tot}}=\pm\frac{1}{2}$). Given $\frac{I_{(0,0,1.5)}(7.5~\mathrm{T})}{I_{(0,0,1.5)}(0~T)}=0.87$ from our data in Fig.~\ref{elastic}~(b), we estimate a canting, $\mathrm{\phi}\approx 25.5^\circ$ at B=7.5~T, in reasonable agreement $\mathrm{\phi}\approx 27.9^\circ$ estimated from the high field data in Ref.~\citep{Hoffmann2021highfield}. On the other hand, $B_\mathrm{sat}$ is determined to be $\sim$22~T from the quadratic fit to $I_{(0,0,1.5)}(\mathrm{B})$ (black solid line in Fig.~\ref{elastic}~(b)), somewhat bigger than a $B_\mathrm{sat}=16.3~\mathrm{T}$ directly measured in Ref.~\citep{Hoffmann2021highfield}. We note that a precise determination of $B_\mathrm{sat}$ here is probably difficult given the small variation of $I_{(0,0,1.5)}(\mathrm{B})$ in the field range covered in our experiment.

To understand the effect of varying field directions on the spin reorientation in CoTiO$_3$, we remounted the crystal in the (H,H,L)-plane, with a field applied along the (1,-1,0) direction that is 30$^\circ$ from the $\mathbf{b}$-axis. The intensity of the (1,1,1.5) Bragg peak was used to track the moment direction as a function of the field. As shown in Fig.~\ref{elastic}~(b), aside from a different high field intensity due to a different projection of $\mathbf{Q}$ in the $\mathbf{ab}$ plane, field dependence at (1,1,1.5) is almost identical to that at (1,0,0.5). More detailed angle dependence was obtained by carrying out magnetization measurements for different field directions as shown in Fig.~\ref{elastic}~(a), where we observed identical $M(\mathrm{B})$ for all field directions. Our results here are consistent with a fine angle dependence reported in Ref.~\citep{dey2021magnetostriction} showing a variation of at most 3$\%$ at 1~T. The lack of angle dependence observed here is quite puzzling: depending on whether the field is along the easy or hard axis determined by the intrinsic spin anisotropy in CoTiO$_3$, one should naively expect a difference in the field dependence as has been observed in other simple collinear magnets\citep{spinfloptheoryPRB2007}. As we will discuss in Section~\ref{ZCmagnondiscussion}, this is naturally explained by the unusual domain structure at zero field.
\FloatBarrier

\subsection{Zone Center Magnon}\label{2A}
\subsubsection{Temperature Dependence}
As discussed in the last subsection, the in-plane spin reorientation transition implies the existence of a finite spin anisotropy that prevents the ordered moments from freely rotating in CoTiO$_3$ at zero field. The same spin anisotropy opens a gap in the Goldstone mode at the magnetic zone center (ZC). The ZC gap was not resolved in our first INS measurement\citep{BoYuan2020PRX}, but was determined to be $\sim$~1~meV by a subsequent high-resolution INS study by Elliot \textit{et al.}~\citep{Elliot2021}, which was later confirmed by both terahertz and Raman spectroscopy measurements\citep{li2022ringexchange}. Assuming the reported high symmetry crystal structure to be correct, this gap is forbidden on a mean-field level when only bilinear spin interactions are considered. Quantum order by disorder\citep{Elliot2021}, which corrects the mean-field energy by considering zero-point fluctuations of the high energy magnons, and a sixth-order ring exchange interaction\citep{li2022ringexchange} were proposed as possible mechanisms to account for observed ZC magnon gap.

\begin{figure}[hbt]
\includegraphics[width=0.5\textwidth]{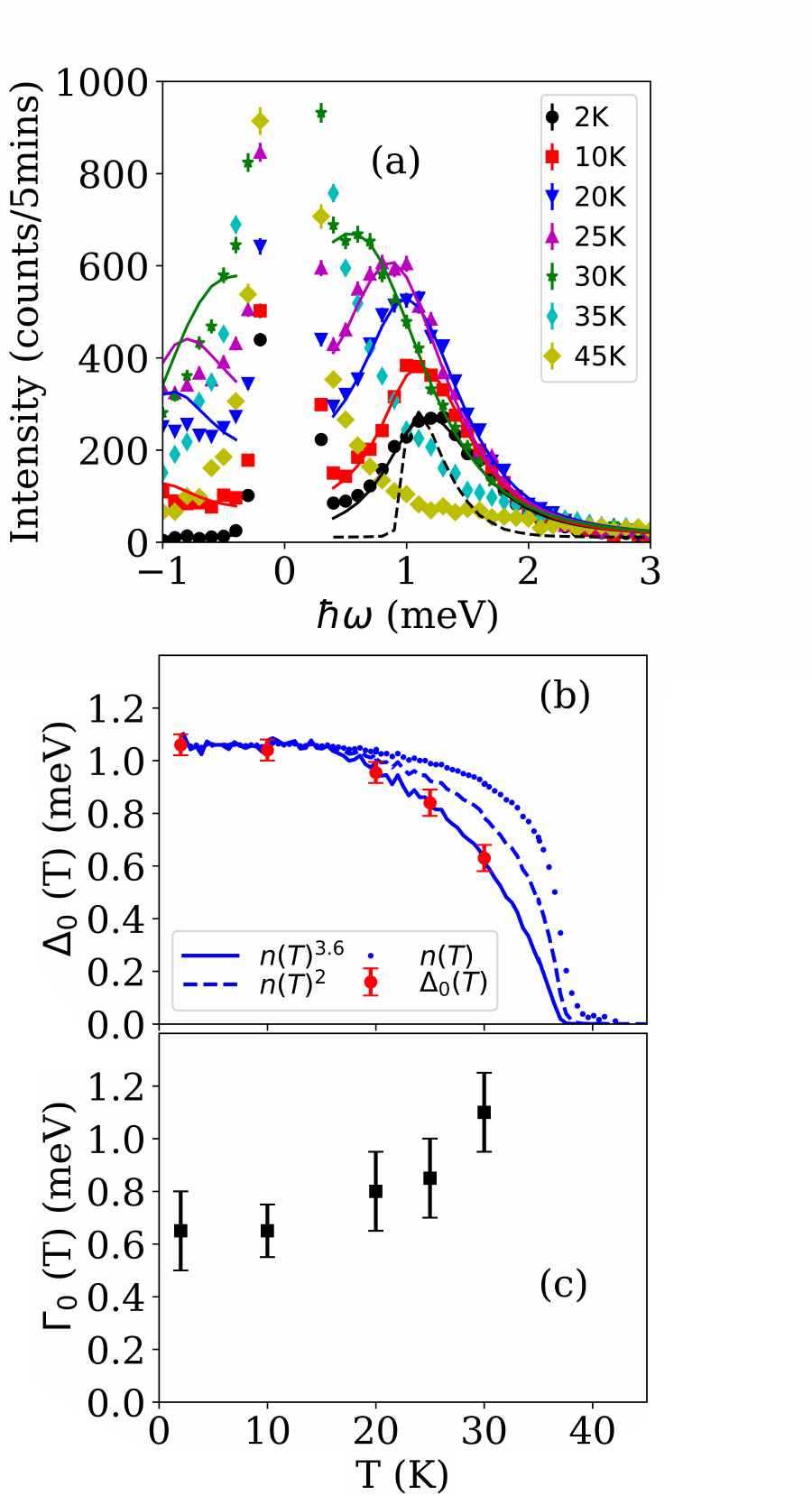}
\caption{\label{ZCtempdep} (a) Constant $\mathbf{Q}$ scans at the magnetic zone center, (0,0,1.5), at different temperatures below and above $\mathrm{T_N}=38$~K measured at the SPINS triple axis spectrometer using fixed final energy, $\mathrm{E_f}=$5~meV. Solid lines are fit to a damped harmonic oscillator model convolved with the SPINS instrumental resolution as described in the text. The dashed line is the expected line shape at 2~K for a resolution-limited magnon peak. (b, c) Temperature dependence of (b) the energy, $\Delta_0~(\mathrm{T})$, and (c) the width, $\Gamma_0~(\mathrm{T})$, of the zone center magnon peak extracted from the fit in (a). Blue lines in (b) are temperature dependence of different powers of the measured order parameter, $n(\mathrm{T})$, whose base temperature value has been set to that of $\Delta_0$ for comparison. $n(\mathrm{T})$ shown here was obtained experimentally by taking the square root of the Bragg peak intensity at (0,0,1.5) measured at SPINS.}
\end{figure}

Using cold neutron triple-axis measurements, we confirmed the previously determined gap size by observing a magnon peak at $\sim$~1~meV in a constant $\mathbf{Q}$ scan at the magnetic ZC as shown in Fig.~\ref{ZCtempdep}~(a). Further measurements were carried out at higher temperatures to study the evolution of the ZC magnon gap as magnetic order gets suppressed when T approaches $\mathrm{T_N}$. A clear softening and damping of the ZC magnon peak is observed as temperature increases, consistent with the typical behaviours observed in an ordered magnet. We fit the inelastic part of the spectrum ($|\hbar\omega|~\geq~0.4$~meV) to a damped harmonic oscillator model (DHM) convolved with the instrumental resolution. The DHM is defined as,
\begin{equation}
\label{Eq1}
\begin{aligned}
S(\mathbf{Q},\hbar\omega,T)=&\frac{1}{\Omega_\mathbf{Q}}\frac{A(\mathrm{T})}{1-e^{-\frac{\hbar\omega}{\mathrm{k_B T}}}}\bigg( \frac{\frac{1}{2}\Gamma_0(\mathrm{T})}{(\hbar\omega-\Omega_\mathbf{Q})^2+\left[\frac{1}{2}\Gamma_0(\mathrm{T})\right]^2}\\
&-\frac{\frac{1}{2}\Gamma_0(\mathrm{T})}{(\hbar\omega+\Omega_\mathbf{Q})^2+\left[\frac{1}{2}\Gamma_0(\mathrm{T})\right]^2} \bigg),
\end{aligned}
\end{equation}
where the phenomenological magnon dispersion near the ZC, $\Omega_\mathbf{Q}$, is taken to be $\Omega_\mathbf{Q}~=~\sqrt{(v_\parallel Q_\parallel)^2+(v_\perp Q_\perp)^2+\Delta_0(\mathrm{T})^2}$. In Eq.~\ref{Eq1}, the overall intensity, $A(\mathrm{T})$, the ZC magnon gap, $\Delta_0(\mathrm{T})$, and the width, $\Gamma_0(\mathrm{T})$, are treated as temperature-dependent fitting parameters, whereas the in-plane and out-of-plane spin wave velocities, $v_\parallel$ and $v_\perp$, are taken to be their base temperature values (See Supplemental Materials). The use of temperature independent spin wave velocities can be justified by a lack of significant temperature dependence of the magnon spectrum up to 30~K from our previous low resolution measurement (see Supplemental Materials). Fit to Eq.~\ref{Eq1} is only performed for T$\leq$30~K where the peaks in the neutron energy gain and loss side are clearly separable. 

Temperature dependence of $\Delta_0(\mathrm{T})$ and $\Gamma_0(\mathrm{T})$ are plotted in Fig.~\ref{ZCtempdep}~(b) and (c), respectively, clearly showing a softening and damping of the ZC magnon with increasing temperature. A quantity of particular interest that can be extracted from Fig.~\ref{ZCtempdep}~(b) is the scaling of $\Delta_0(\mathrm{T})$ with the order parameter, $n(\mathrm{T})\equiv|\mathbf{n}|$, the latter of which is determined experimentally by taking the square root of the measured (0,0,1.5) Bragg peak intensity. Interestingly, by comparing the temperature dependence of different powers of the measured order parameter, $n(\mathrm{T})^\alpha$, with $\Delta_0(\mathrm{T})$ as shown in Fig.~\ref{ZCtempdep}~(b), we found neither the order parameter, $n(\mathrm{T})$, nor the order parameter squared (equivalent to the magnetic Bragg peak intensity), $n(\mathrm{T})^2$, describes the temperature dependence of the ZC gap. Instead, it scales as a relatively high power of $n(\mathrm{T})$ as $\Delta_0(\mathrm{T})~\propto~n(\mathrm{T})^\alpha$, with $\alpha=3.6\pm 1.0$. At T~=~0, the ZC gap due to quantum order by disorder and the ring exchange should scale as $n^\frac{1}{2}$ \citep{rauObD2018}, and $n^5$, respectively. Assuming the same scaling holds at finite temperature (likely to be true at least within a random phase approximation), the large $\alpha$ determined here therefore appears to be consistent with a ZC gap opened by the ring exchange interaction. However, a more careful measurement of $\Delta_0(\mathrm{T})$ is required to confirm this. Aside from more temperature points for a robust determination of $\alpha$, a more appropriate functional form than Eq.~\ref{Eq1} is required to describe the data at high temperatures. Strictly speaking, Eq.~\ref{Eq1} starts to break down when $\Gamma_0~\sim~\Delta_0$, as evidenced by the poor fit on the neutron energy gain side at $T~\geq~20$~K, and a critical scattering form of $S(\mathbf{Q},\hbar\omega,\mathrm{T})$ \citep{MnF2criticalscattering, K2CuF4criticalscattering} is probably needed for a better fit as $T\rightarrow~\mathrm{T_N}$.

\subsubsection{Field Dependence}\label{section2A1}
A remarkable feature from the above analysis is a significant broadening of the ZC magnon peak already at base temperature, which is evident by comparing the measured lineshape (black circles in Fig.~\ref{ZCtempdep}~(a)) and a simulated one with $\Gamma_0=0$ (dashed line in Fig.~\ref{ZCtempdep}~(a)). To understand the origin of the observed broadening, we studied the evolution of the ZC magnon across the in-plane spin reorientation transition driven by an in-plane magnetic field. 

\begin{figure*}[hbt!]
\includegraphics[width=1\textwidth, trim=0cm 0cm 0cm 0cm, clip]{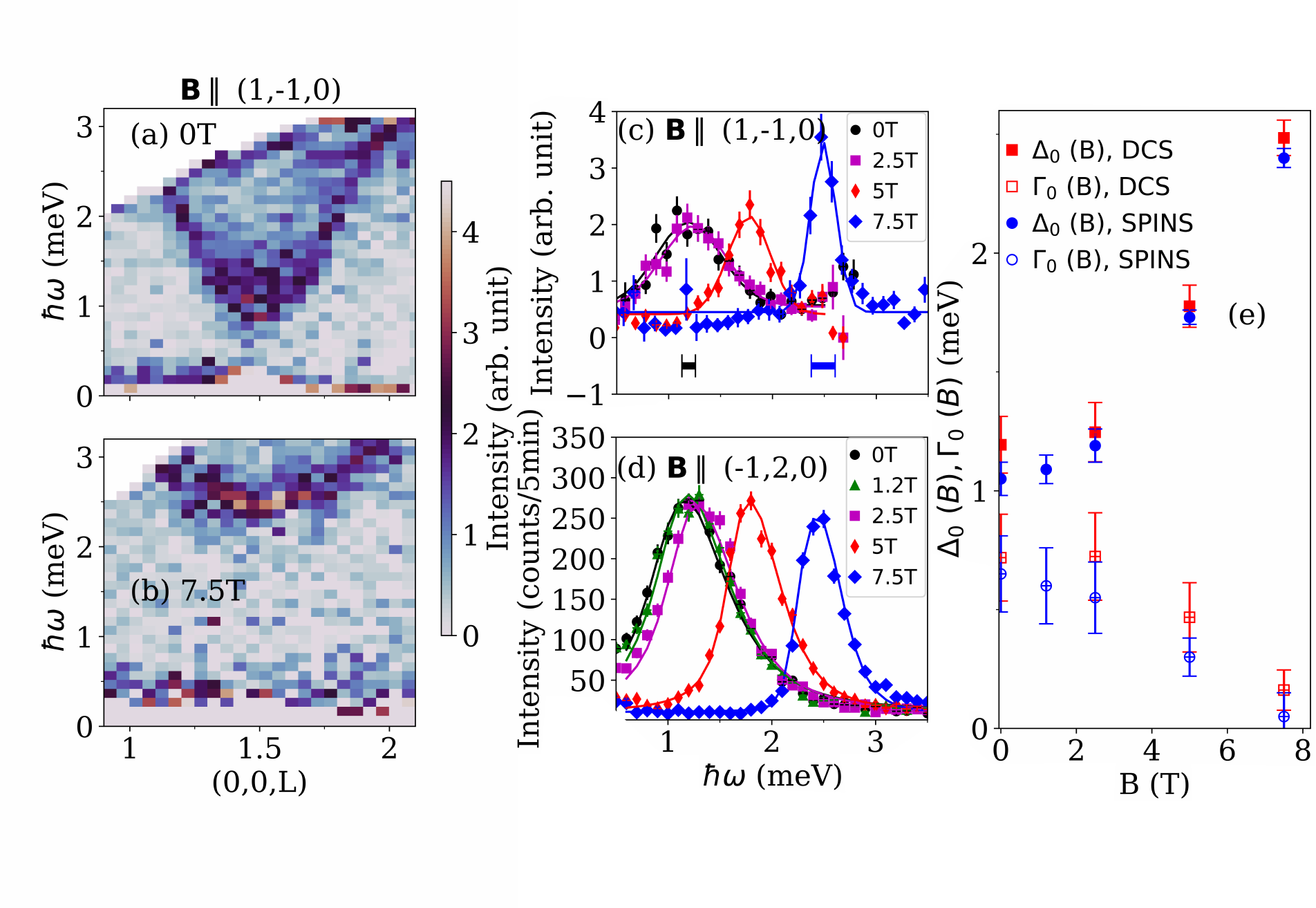}
\caption{\label{ZCfielddep} (a-b) Magnon spectrum near the magnetic zone center (ZC), (0,0,1.5), (a) in the absence of a magnetic field and (b) with a 7.5~T field applied along the (1,-1,0) direction. (c) Constant $\mathbf{Q}$ cut of the ZC magnon spectrum for different field strengths in the (1,-1,0) direction. An integration range of $\pm 0.1$~rlu and $\pm 0.02$~rlu were used along the L and the $(\zeta,\zeta,0)$ directions around the magnetic ZC at (0,0,1.5), corresponding to $\sim~\pm0.05\AA^{-1}$ in both directions. (Effect of integration ranges on the lineshape of the constant $\mathbf{Q}$ cuts is shown in the Supplemental Materials.)  Solid lines are obtained by fitting the data to a Gaussian on top of a constant background. Data in (a-c) were obtained at the DCS TOF spectrometer with the CoTiO$_3$ crystal mounted in the (H,H,L) plane in a vertical field magnet. Measurements at 0~T,2.5~T and 5~T used an incident energy, E$_i$, of 4.04~meV, whereas that at 7.5T used an E$_i$ of 5.66~meV. Instrumental resolution for these two E$_i$'s are shown as horizontal bars at the bottom. (d) Constant $\mathbf{Q}$ scan at (0,0,1.5) with different fields applied along the (-1,2,0) direction. The data were obtained at the SPINS triple-axis spectrometer using a fixed, $\mathrm{E_f}=$5~meV, by mounting the sample in the (H,0,L) plane in a vertical field magnet. The solid lines are fit to the data after convolving with the SPINS resolution function as described in the text.  (e) Field dependence of the energy, $\Delta_0~(\mathrm{B})$ (solid symbols), and width, $\Gamma_0~(\mathrm{B})$ (open symbols), of the ZC magnon obtained from the fit to the DCS and the SPINS data in (c) and (d), respectively. The parameters extracted from the DCS data are shown as red squares, while those extracted from the SPINS data are shown as blue circles. For the fit of the DCS data in (c), The peak position of the Gaussian was taken to be $\Delta_0$. To account for the slightly different energy resolution of the 7.5~T data from the rest, intrinsic width of the magnon peak, $\Gamma_0$, was determined by removing the instrument resolution ($\Gamma_\text{res}$) from the full-width at half maximum extracted from the fit ($\Gamma_\text{measured}$) or, $\Gamma_0=\sqrt{\Gamma_\text{measured}^2-\Gamma_\text{res}^2}$.}
\end{figure*}

Effects of an in-plane magnetic field on the ZC magnon are most clearly seen in Fig.~\ref{ZCfielddep}~(a), where we compare the magnon spectra near (0,0,1.5) measured by cold neutron time-of-flight (TOF) spectroscopy at 0~T and 7.5~T for a field along (1,-1,0). In addition to an increase in the ZC magnon gap due to the Zeeman interaction, the width of the ZC magnon shows a clear reduction at high field compared to zero field. The observed width change is quite striking especially considering that a larger incident energy is actually used at 7.5~T than at 0~T which offers poorer energy resolution. To extract the field dependence of the energy, $\Delta_0~(\mathrm{B})$, and the width, $\Gamma_0~(\mathrm{B})$, we made constant $\mathbf{Q}$ cuts of the TOF data in Fig.~\ref{ZCfielddep}~(a-b). The results shown Fig.~\ref{ZCfielddep}~(c) reveal a clear sharpening of the magnon mode, which becomes resolution limited at 7.5~T [the horizontal bar in Fig.~\ref{ZCfielddep}~(c) indicates the instrumental energy resolution]. As we show in the Supplemental Materials, the width of the constant $\mathbf{Q}$ cut at 0~T is independent of the choice of integration range along $(\zeta,\zeta,0)$ and L, thus ruling out any artificial broadening due to the finite integration range used in producing these cuts. $\Delta_0~(\mathrm{B})$ and $\Gamma_0~(\mathrm{B})$ determined by fitting the constant $\mathbf{Q}$ cut of the TOF data in Fig.~\ref{ZCfielddep}~(c) to a Gaussian is shown in Fig.~\ref{ZCfielddep}~(e) as red solid and open squares, respectively. Consistent with the expected behaviour for an ordered antiferromagnet undergoing a spin reorientation transition, $\Delta_0$ stays roughly constant at small fields and shows an almost linear increase at high fields, signaling a transition from a low-field regime dominated by the intrinsic spin anisotropy of the system to a high field regime where the Zeeman coupling to the applied field becomes important. On the other hand, $\Gamma_0 (\mathrm{B})$ exhibits B-dependence behavior almost exactly opposite to that of $\Delta_0 (\mathrm{B})$ and shows a clear decrease in the high-field state.

A similar field-dependent study of the ZC magnon for $\mathbf{B}\parallel (-1,2,0)$ was carried out at the cold neutron triple-axis spectrometer at SPINS. Constant $\mathbf{Q}$ scans at (0,0,1.5) shown in Fig.~\ref{ZCfielddep}~(d) for different fields along (-1,2,0) look almost identical to the TOF data in Fig.~\ref{ZCfielddep}~(c). On the other hand, the high data quality in Fig.~\ref{ZCfielddep}~(d) and a better known instrumental resolution function for SPINS allowed a more precise determination of $\Gamma_0 (\mathrm{B})$ and $\Delta_0 (\mathrm{B})$. As shown by solid lines in Fig.~\ref{ZCfielddep}~(d), we fit the SPINS data by convolving Eq.~\ref{Eq1} with the same phenomenological dispersion used to fit the temperature-dependent data and the instrumental resolution. When fitting the data at each field, the spin wave velocities are fixed to the values determined from the TOF data (see Supplemental Materials), while  $\Gamma_0 (\mathrm{B})$ and $\Delta_0 (\mathrm{B})$ are allowed to vary. As shown by the fitting results in Fig.~\ref{ZCfielddep}~(e), there is a clear quantitative agreement between the triple-axis and the TOF data, obtained with magnetic fields along two in-plane directions unrelated by symmetry. Combining the elastic and inelastic results in Fig.~\ref{elastic} and Fig.~\ref{ZCfielddep} respectively shows that both the static as well as the low energy dynamical properties across the spin reorientation transition are independent of field direction in CoTiO$_3$. 
\FloatBarrier
 
\subsection{Dirac Magnon Near the $\mathbf{K}$ Point}\label{sec2B}
\subsubsection{Zero field}\label{sec2B1}
\begin{figure}[ht!]
\includegraphics[width=0.5\textwidth]{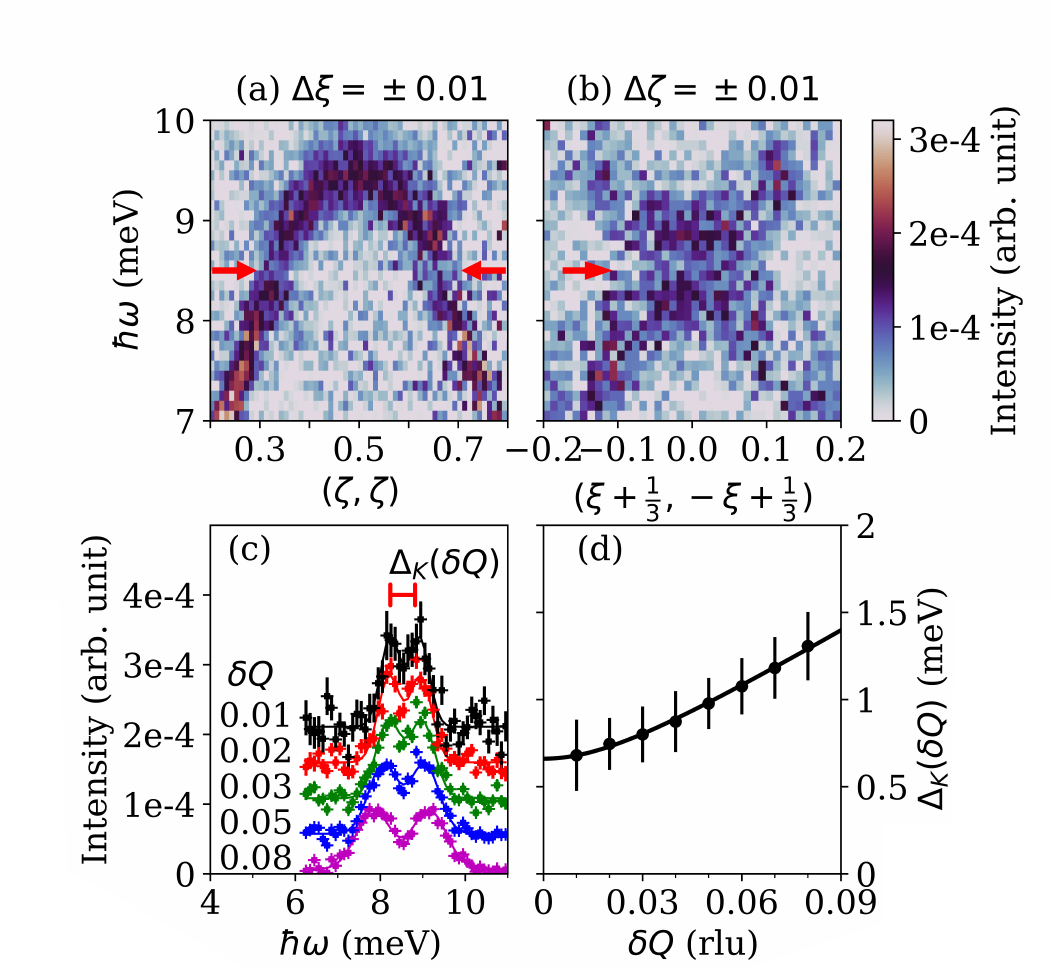}
% Here is how to import EPS art
\caption{\label{Dirac0T} (a-b) Magnon spectrum near the $\mathbf{K}=(\frac{1}{3},\frac{1}{3})$ point along the (a) $(\zeta,\zeta)$ and (b) $(\xi,-\xi)$ directions measured at the SEQUOIA TOF spectrometer using an incident energy, $\mathrm{E_i}$, of 12~meV. The data within $\pm$~0.01~rlu transverse to the directions shown in (a) and (b) were integrated to produce these plots. Due to the lack of $L$ dependence in the magnon spectrum near $\mathbf{K}$, the data were integrated over all $L$ probed in the experiment, $-4\lesssim L \lesssim 4$ to improve the signal to noise ratio. (c) Constant $\mathbf{Q}$ cut at the $\mathbf{K}$ point for different integration range, $\delta \mathrm{Q}$. For each $\delta \mathrm{Q}$ (in rlu), intensity within the rectangular box defined by $-\delta \mathrm{Q} +\frac{1}{3}<\zeta<\delta \mathrm{Q} +\frac{1}{3}$ and $-\delta \mathrm{Q}<\xi<\delta \mathrm{Q}$ were integrated and plotted as a function energy transfer. $\delta \mathrm{Q}$=0.01 in rlu corresponds to an integration range of $\pm~0.025~\AA^{-1}$ and $\pm~0.014~\AA^{-1}$ along $(\zeta,\zeta)$ and $(\xi,-\xi)$, respectively. Constant $\mathbf{Q}$ cuts using different $\delta \mathrm{Q}$ are vertically offset for clarity. Solid lines are fit to these cuts using two Gaussians having equal width and height. (d) Splitting of the two peaks in (c), $\Delta_\mathrm{K}$, as a function integration range, $\delta \mathrm{Q}$.}
\end{figure}

\begin{figure*}[hbt]
\includegraphics[width=1\textwidth]{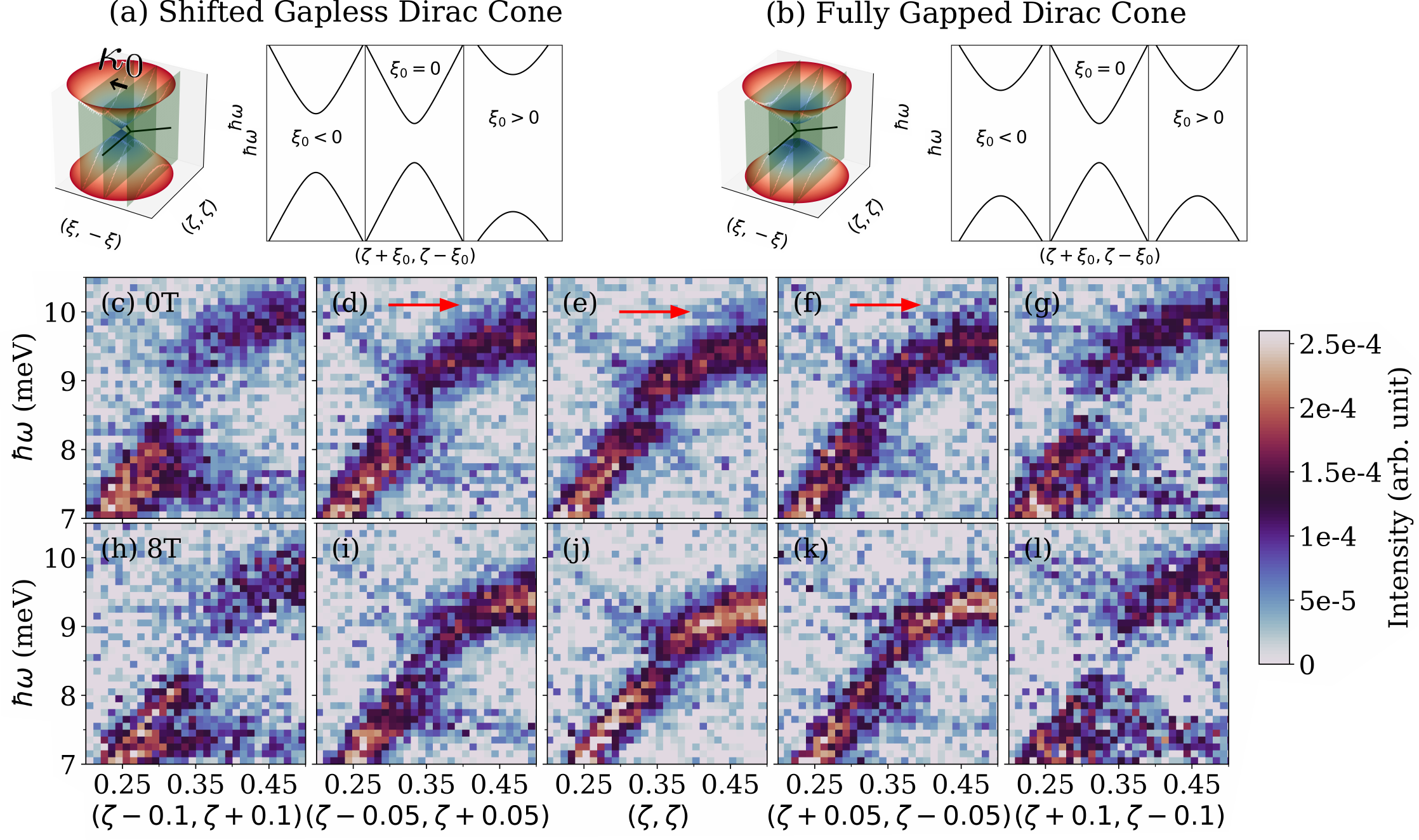}
\caption{\label{Diracfielddep} (a-b) Schematics comparing the magnon dispersions near the $\mathbf{K}$ point for a (a) gapless Dirac cone shifted away from $\mathbf{K}$ by $\kappa_0$  in the $(\xi,-\xi)$ direction and (b) fully gapped Dirac cone centered at $\mathbf{K}$ (c-g) Magnon spectra in the vicinity of the $\mathbf{K}$ point in zero magnetic field. Momentum directions shown here are along $(\zeta+\xi_0,\zeta-\xi_0)$ for fixed $\xi_0$'s, corresponding to the momentum directions in the slices shown schematically in (a-b). An integration range of $\pm$0.02 rlu in the $(\xi,-\xi)$ direction was used, which was chosen to improve the signal to noise ratio without introducing distortion to the spectrum associated with a large integration range [See Fig.~\ref{Dirac0T}(c) and Fig.~\ref{Diraccut}]. Red horizontal arrows denote the weak intensity near the main magnon mode that gives rise to the shoulder in the constant $\mathbf{Q}$ cuts in Fig.~\ref{ZBcut}. (h-l) are the same as (c-g) but were obtained in the presence of an 8~T field along the (1,-1,0) direction. Data in (c-l) were obtained at SEQUOIA using an $\mathrm{E_i}=$12~meV.}
\end{figure*}

As pointed out by Elliot \textit{et\,al.}\citep{Elliot2021}, a key manifestation of the non-trivial exchange anisotropy in CoTiO$_3$ is the existence of a gap in the magnon dispersion at the $\mathbf{K}$ point, which would, in a simple XXZ model, host a Dirac-cone like dispersion with linear crossing. This feature was unresolved in our previous measurements with a coarse $\sim$1~meV energy resolution \citep{BoYuan2020PRX}, but could be resolved by using a lower incident energy, E$_i$, of 12~meV offering a much better energy resolution of 0.27~meV (at the elastic line). High resolution magnon spectra measured at the SEQUOIA TOF spectrometer along the $(\zeta,\zeta)$ and $(\eta+\frac{1}{3},-\eta+\frac{1}{3})$ directions are shown in Fig.~\ref{Dirac0T}~(a) and (b), respectively. [Due to the lack of L-dependence of the magnon at $\hbar\omega\geq 7$~meV (see Supplemental Materials), the data shown in Fig.~\ref{Dirac0T}-~Fig.~\ref{ZBcut} have been integrated over all L to improve signal to noise ratio. The $\mathbf{Q}$ transfers shown in these figures are projections onto the \textbf{ab} plane.] A clear depletion of the magnon intensity, as highlighted by the red arrows in Fig.~\ref{Dirac0T}~(a,b), is observed near the $\mathbf{K}$ point at $(\frac{1}{3},\frac{1}{3})$ and $(\frac{2}{3},\frac{2}{3})$, consistent with the data in Ref~\citep{Elliot2021}. This is further illustrated by the constant $\mathbf{Q}$ cut at the $\mathbf{K}$ point in Fig.~\ref{Dirac0T}~(c), where the integrated neutron intensity in a small rectangular box centered at $\mathbf{K}$ was plotted as function of energy transfer, $\hbar\omega$. A splitting of the magnon peak of $\sim 1$~meV was observed in the constant $\mathbf{Q}$ cut, confirming the existence of a gap between the optical and the acoustic magnon branches at the $\mathbf{K}$ point. To rule out any artifacts caused by using a large in-plane integration range when producing the constant $\mathbf{Q}$ cut \citep{CrCl3Diracmagnon2022,CrXTe3scienceadvance2021, CrBr3tempdep2022, CrBr3DiracgapPRB2021}, we compared cuts made with different integration range, $\delta \mathrm{Q}$, in Fig.~\ref{Dirac0T}(c). When a large $\delta \mathrm{Q}$ is used, the apparent width, and more importantly the apparent splitting between the two magnon peaks increase, leading to an overestimation of the true magnon gap at the $\mathbf{K}$ point in the study of CrBr$_3$\citep{CrBr3tempdep2022, CrBr3DiracgapPRB2021} and CrCl$_3$\citep{CrCl3Diracmagnon2022}. In CoTiO$_3$, we found the constant $\mathbf{Q}$ cuts shown in Fig.~\ref{Dirac0T}(c) to be almost unchanged when using a $\delta \mathrm{Q}\lesssim 0.03$, clearly demonstrating the robustness of the gap at the $\mathbf{K}$ point. The splitting between the two magnon peaks, $\Delta_\mathrm{K} (\delta \mathrm{Q})$, as a function of $\delta \mathrm{Q}$ obtained by fitting the constant $\mathbf{Q}$ cuts to two Gaussians is shown in Fig.~\ref{Dirac0T}~(d). Clearly, $\Delta_\mathrm{K} (\delta \mathrm{Q})$ plateaus to a non-zero value, $\Delta_{\mathrm{K}}$, in the limit of $\delta \mathrm{Q}=0$, representing the true gap size at the $\mathbf{K}$ point. By fitting the data in Fig.~\ref{Dirac0T}(d) to $\Delta_\mathrm{K} (\delta \mathrm{Q})=\frac{1}{2}\Delta_\mathrm{K}+\sqrt{\frac{\Delta_\mathrm{K}^2}{4}+(c\delta \mathrm{Q})^2}$ where $c$ denotes the magnon velocity near $\mathbf{K}$(see Supplemental Materials for the derivation of this expression), we determined $\Delta_\mathrm{K}$ to be 0.7(2)~meV, in agreement with the value reported by Elliot \textit{et\,al} \citep{Elliot2021}.

Understanding the origin of the $\mathbf{K}$ point gap has posed a major theoretical challenge in the study of CoTiO$_3$. Importantly, the presence of an inversion center between two adjacent Co$^{2+}$ layers combined with the simple collinear antiferromagnetic order preserves the so-called ``magnetic inversion'' symmetry of the Bogoliubov Hamiltonian that protects the Dirac nodal line in this material\citep{Elliot2021,ArunKreinPRB2022}. This led to the proposal of a shifted Dirac cone scenario shown schematically in Fig.~\ref{Diracfielddep}~(a), which could arise in the presence of bond-dependent interaction in CoTiO$_3$\citep{Elliot2021}. In Fig.~\ref{Diracfielddep}~(a), we show a case where the Dirac cone remains gapless but is slightly shifted away from the $\mathbf{K}$ point by $\kappa_0$ in the (1,-1) direction. Consequently, a gap given by $2c\kappa_0$ appears at the $\mathbf{K}$ point. This is contrasted with the scenario of a fully gapped Dirac cone centered at $\mathbf{K}$ shown in Fig.~\ref{Diracfielddep}~(b). Although the magnon dispersion along $(\zeta,\zeta)$ exactly passing through $\mathbf{K}$ looks identical in these two scenarios, they can be distinguished by examining the magnon dispersions along $(\zeta+\xi_0,\zeta-\xi_0)$ with a non-zero $\xi_0$. Since the gap closes at $(\kappa_0,-\kappa_0)$ in the first scenario, one should observe a gapless crossing along the $(\zeta+\xi_0,\zeta-\xi_0)$ direction with $\xi_0$~=~$\kappa_0$. In addition, since the Dirac cone is now displaced to the left of the $\mathbf{K}$ point, the magnon spectrum along $(\zeta+\xi_0,\zeta-\xi_0)$ should show a large asymmetry for $\pm \xi_0$ as shown by the simulated magnon dispersions in Fig.~\ref{Diracfielddep}~(a). 

Although there is a clear distinction between the two scenarios in a mono-domain state with a well-defined ordered moment direction, attempting to model the zero-field spectrum using linear spin wave theory is complicated by the presence of both structural and magnetic domains at 0~T, each with a distinct excitation spectrum. The presence of two structural domains related by a two-fold rotation around (1,1,0) has been inferred from the relative intensities of two inequivalent Bragg peaks related by the two-fold rotation (see Supplemental Materials). Its effect is to symmetrize the magnon spectrum along $(\zeta \pm \xi_0,\zeta\mp \xi_0)$, independent of the spin model. On the other hand, the configuration of magnetic domains at 0~T is entirely unknown in CoTiO$_3$. A simple configuration with discrete magnetic domains related by the $\mathcal{C}_3$ and the time-reversal symmetry was assumed by Elliot \textit{et\,al} \citep{Elliot2021} when modelling the zero-field spectrum. However, as we discuss in Section~\ref{ZCmagnondiscussion}, a more complicated situation with extended domain walls is likely present in CoTiO$_3$ at zero field.  

Furthermore, the zero-field ordered moment directions, necessary for any spin wave modelling, could not be experimentally determined in CoTiO$_3$. Elliot \textit{et\,al} \citep{Elliot2021} addressed this difficulty by assuming an ordered moment direction determined by a quantum order-by-disorder (ObD) mechanism using the same form of bond-dependent interaction that also displaces the Dirac cone. However, this assumption is questioned by a recent optical study\citep{li2022ringexchange} which argued that the ring exchange interaction, instead of quantum ObD, is responsible for the observed ZC gap

\subsubsection{8~T Magnetic Field}
To address the above difficulties, we carried out high resolution TOF measurement with an 8~T field applied along (1,-1,0). The crystal was mounted in the (H,H,L) plane to access the $\mathbf{K}$ point at $(\frac{1}{3},\frac{1}{3})$. As we showed in Section~\ref{spinflop}, an 8~T field along (1,-1,0) is sufficient to drive a spin reorientation transition in CoTiO$_3$ from a zero-field multi-domain state to a high field mono-domain state with a \textit{known} canted magnetic structure. This resolves the ambiguities in modelling the spin wave spectra at zero field where both the ordered moment directions and magnetic domain configuration are unknown. The magnon spectra near the $\mathbf{K}$ point in the high-field state are shown in Fig.~\ref{Diracfielddep}~(h-l). By comparing with the zero-field spectra shown in Fig.~\ref{Diracfielddep}~(c-g), we arrive at the important observation that the magnon spectra at 8~T are almost unchanged compared to those at 0~T. As shown by Elliot \textit{et\,al} \citep{Elliot2021} and our systematic investigation using linear spin wave theory (See Supplemental Materials), the shift of the Dirac cone near the $\mathbf{K}$ point with bond-dependent interactions is strongly dependent on the direction of ordered moments. Given that there is a clear change of the ordered moment directions across the spin reorientation transition, it is difficult to explain the observation of almost identical magnon spectra near the $\mathbf{K}$ point at 0~T and 8~T by a shifted Dirac cone scenario with bond-dependent interactions.  

\begin{figure}[hbt]
\includegraphics[width=0.5\textwidth]{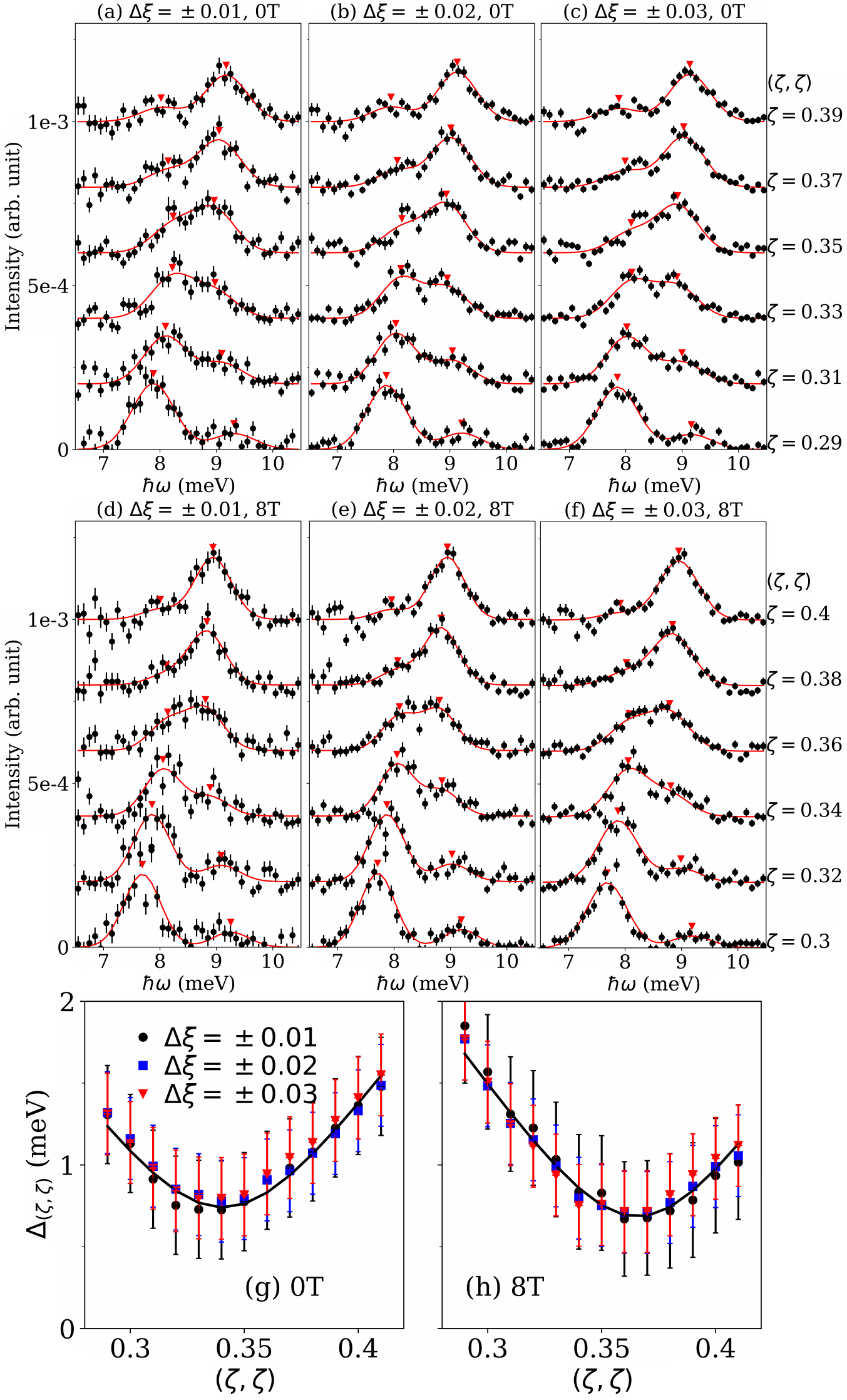}
\caption{\label{Diraccut}(a-c) Constant $\mathbf{Q}$ cuts of the zero field magnon spectrum near $\mathbf{K}=(\frac{1}{3},\frac{1}{3})$ at various points along the $(\zeta,\zeta)$ direction. The same integration range ($\pm0.01$~rlu) in the $(\zeta,\zeta)$ direction was used in all cuts. Different integration ranges along $(\xi,-\xi)$ used in (a-c) are given in their respective titles. Solid lines are fit to two Gaussians. Widths of the two peaks are fixed for all $(\zeta,\zeta)$'s in each panel to a value determined at a $(\zeta,\zeta)$ point just far enough away from the $\mathbf{K}$ point where the two peaks due to the acoustic and optical magnon branches are clearly separated. (d-f) are the same as (a-c) but obtained for the magnon spectra with an 8~T field applied along (1,-1,0). (g,h) Splitting between the acoustic and optical magnon modes along $(\zeta,\zeta)$, $\Delta_{(\zeta,\zeta)}$, at (g) 0~T and (h) 8~T. The size of the splitting is determined from the separation between the two peaks in (a-f) indicated by the downward triangles. The solid lines in (g) and (f) are fit of the measured splitting along $(\zeta,\zeta)$ to $\Delta_{(\zeta,\zeta)}=\sqrt{\Delta_K^2+[2c(\zeta-\zeta_0)]^2}$. In this expression, $\Delta_K$ is the size of the Dirac gap and $c$ is the magnon velocities near the $\mathbf{K}$ point. Since the Dirac cone is no longer centered exactly at the $\mathbf{K}$ point at 8~T, we also allowed its location $\zeta_0$ to vary in the fit. $\zeta_0$ is $\frac{1}{3}$ and $\sim0.37$ at 0T and 8T, respectively.}
\end{figure}

To look for any small difference in the size of the Dirac gap at 0~T and 8~T, we made constant $\mathbf{Q}$ cuts of the spectrum in Fig.~\ref{Diracfielddep}~(e) and (j) at different points along $(\zeta,\zeta)$. The results are shown in Fig.~\ref{Diraccut}~(a-c) and Fig.~\ref{Diraccut}~(d-f), respectively. For each field, cuts made with different ranges of integration in the $(\xi,-\xi)$ direction are presented to check the robustness of the analyses. Lineshapes at different $(\zeta,\zeta)$'s show qualitatively similar behaviours in all plots: at $\mathbf{Q}$ sufficiently far from the $\mathbf{K}$ point, two distinct peaks due to the acoustic and optical magnon branches can be resolved. The two magnon modes approach each other and then separate moving past the $\mathbf{K}$ point. To determine the gap at the $\mathbf{K}$ point, the constant $\mathbf{Q}$ cuts at different $(\zeta,\zeta)$ points in each panel are fit to a sum of two Gaussians with a fixed width. The width is determined by fitting to a cut at a $(\zeta,\zeta)$ point just far enough from $\mathbf{K}$ to have two clearly separated peaks. The splitting between the optical and acoustic branches determined from the fit is shown in Fig.~\ref{Diraccut}~(g) and (h) for 0~T and 8~T, respectively, as a function of the $\mathbf{Q}$ transfer. Aside from a small shift in the $\zeta$ value that minimizes the splitting between the two modes, we found no difference in the size of the Dirac gap at 8~T and 0~T.  As shown by data points with different symbols, this observation is robust for different integration ranges used to obtain the constant $\mathbf{Q}$ cuts.    

\begin{figure}[hbt!]
\includegraphics[width=0.5\textwidth, trim=2cm 0cm 3cm 1cm, clip]{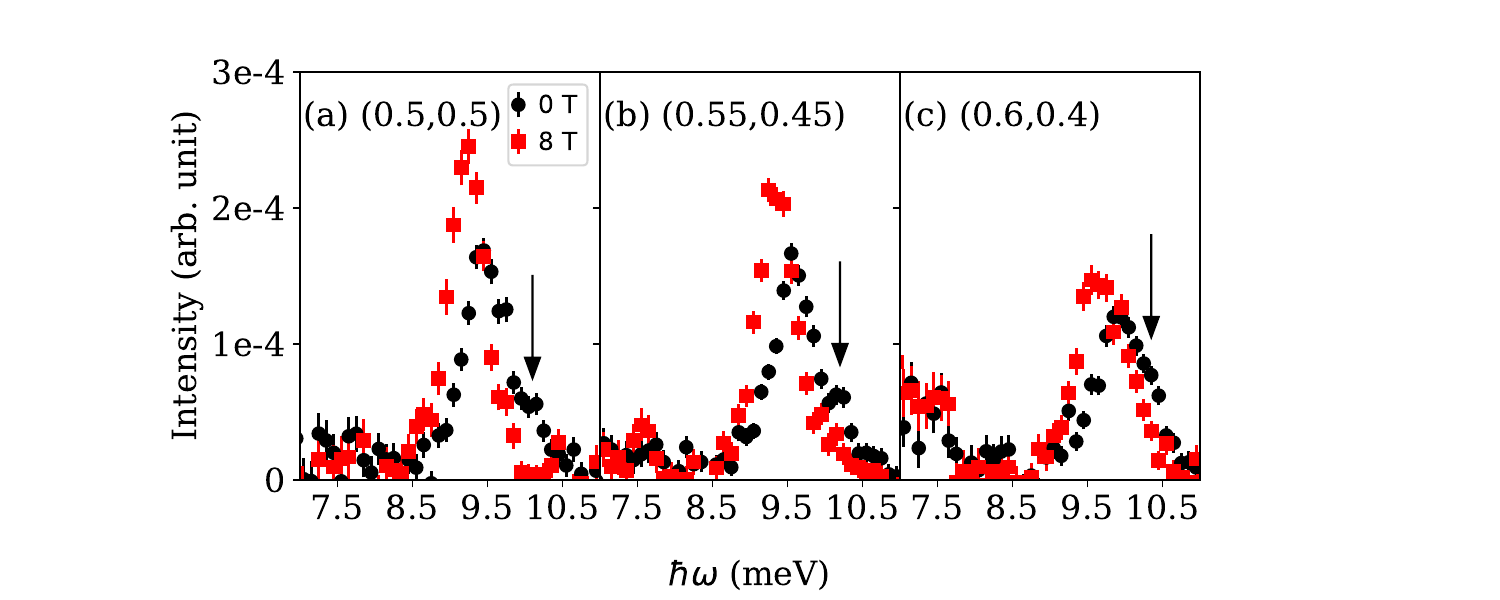}
\caption{\label{ZBcut} Constant $\mathbf{Q}$ cuts of the measured magnon spectra at 0~T and 8~T in Fig.~\ref{Diracfielddep} at (a) $\mathbf{Q}$=(0.5,0.5), (b) $\mathbf{Q}$=(0.55,0.45) and (c) $\mathbf{Q}$=(0.6,0.4). The integration ranges used are $\pm 0.02$~rlu along both the $(\zeta,\zeta)$ and $(\xi,-\xi)$ directions. For (b) and (c), spectra along $(\zeta \pm\xi_0,\zeta \mp\xi_0)$ are first symmetrized to improve the signal to noise ratio before making the cuts.}
\end{figure}

Although the magnon spectra near the $\mathbf{K}$ point shows little change across the spin reorientation transition, subtle differences between the 0~T and 8~T spectra are observed along the Brillouin zone boundary away from the $\mathbf{K}$ point. This is clear by comparing constant $\mathbf{Q}$ cuts of the 0~T and 8~T data at $\mathrm{(0.5+\xi_0,0.5-\xi_0)}$ for different $\xi_0$'s. As shown in Fig.~\ref{ZBcut}, the magnon peak at 8~T shows a slight softening compared to that at 0~T. More interestingly, unlike the constant $\mathbf{Q}$ cuts at 8~T showing a symmetric lineshape, that at 0~T shows a shoulder next to the main magnon peak as indicated by the black vertical arrow. The shoulder in the 0~T originates from a weak mode close to the main magnon mode as indicated by the red horizontal arrows in Fig.~\ref{Diracfielddep}~(d)-(f). The observation that this weak mode is absent in the high-field monodomain state suggests that it originates from either magnetic domains with different ordered moment directions or excitations within the extended domain walls (see Section~\ref{ZCmagnondiscussion}).    
\FloatBarrier

\section{Discussions}
\subsection{Zone Center Magnon}\label{ZCmagnondiscussion}
\begin{figure}[hbt]
\includegraphics[width=0.5\textwidth, trim=0cm 0cm 0cm 0cm, clip]{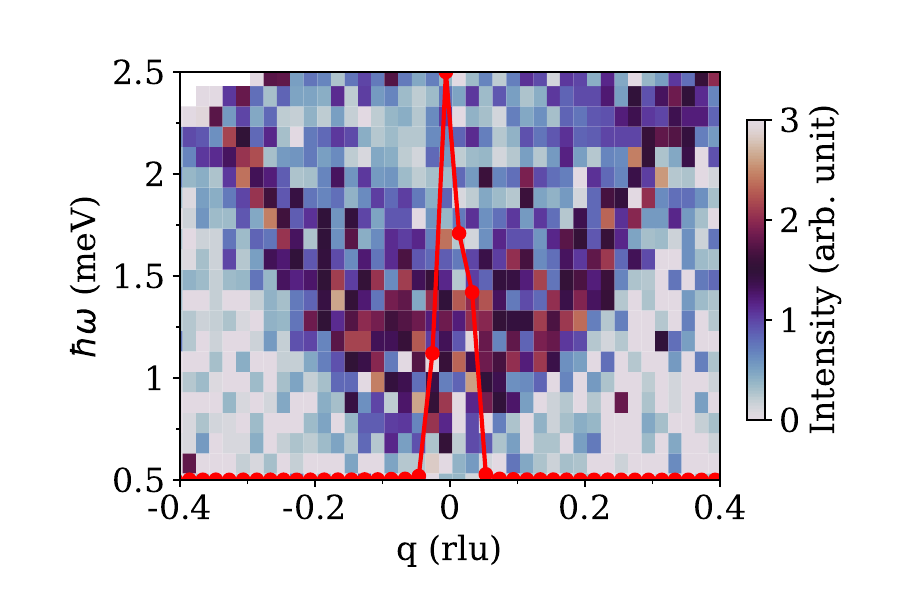}
\caption{\label{ZCmagnonwidth}Zero-field magnon spectrum along L near the magnetic zone center, (0,0,1.5). This is the same data as Fig.~\ref{ZCfielddep}~(a), but plotted in the reduced wave-vector along L defined as, $q$=L-1.5. The peak shown in red is the $q$-scan through the (0,0,1.5) magnetic Bragg peak obtained by integrating the elastic intensity with $|\hbar\omega|<0.1$~meV.}
\end{figure}

A puzzling result from Section~\ref{2A} is the broadening of the ZC magnon at 0~T. Broadening of a magnon mode is usually attributed to damping caused by either magnon-magnon interaction or interaction with other excitations such as phonons. However, this explanation does not apply here as the ZC magnon is kinematically forbidden from decaying, and there is no known low energy phonon mode in CoTiO$_3$. Another possible explanation for damping of the ZC magnon is the finite size effect of magnetic domains at zero field, which could potentially account for the observed width change of the magnon peak across the spin reorientation transition. However, this explanation is unlikely due to the following argument. For a magnetic domain of size $d_0$, only magnon with a wavelength longer than $d_0$, or a reduced wave-vector, $q$, smaller than $\frac{1}{d_0}$ will be damped by scattering with the domain boundary, while those at $q\gtrsim q_0\sim \frac{1}{d_0}$ are unaffected. An upperbound for $q_0$ is given by the width of the magnetic Bragg peak (red data points in Fig.~\ref{ZCmagnonwidth}) to be $\sim$0.05~rlu \footnote{This value is mostly determined by the instrumental resolution. The actual $q_0$ should be much smaller. High resolution synchrotron measurement have been carried out for the doped sample, Co$_{0.65}$Fe$_{0.35}$TiO$_3$\citep{synchrotrondopedCoTiO3}, finding $d_0\sim 2500~\AA$ along both the in-plane and out-of-plane direction equivalent to a $q_0\sim0.005$~rlu.}. However, a close inspection of the ZC magnon spectrum (reproduced in Fig.~\ref{ZCmagnonwidth} together with the q-scan through the (0,0,1.5) magnetic Bragg peak) shows that the magnon is broadened at least up to $q\sim0.2$, clearly outside the width of the Bragg peak. This rules out the finite size effect of magnetic domains as the reason for the observed broadening. 

If not damping, the observed broadening can only be reasonably explained by a distribution of spin anisotropy in the sample, which produces anisotropy gaps of slightly different sizes and manifests as a large width of the ZC magnon with finite energy resolution. Given that the broadening is only observed at 0~T and hence obviously related to the multi-domain state (Fig.~\ref{ZCfielddep}), the most plausible explanation for such a distribution of spin anisotropy is the presence of extended magnetic domain walls. Within these domain walls, the ordered moments are no longer pinned to the easy axes but continuously rotate between two easy directions to smoothly connect two magnetic domains. Consequently, spins within the domain walls experience a slightly different anisotropy field depending on their orientations with respect to the easy axes, resulting in a slightly different ZC magnon gap. In addition to explaining the sharpening of the ZC magnon in the high-field monodomain state, existence of extended magnetic domain walls with a continuous distribution of ordered moment directions also partially restores the in-plane U(1) symmetry and explains the lack of angle dependence observed in Fig.~\ref{elastic}. Physically, existence of extended magnetic domain walls is plausible for two reasons. First, the energy cost for making these domain walls is extremely small as it arises from a very small in-plane spin anisotropy due to either quantum order by disorder \citep{Elliot2021} or ring exchange \citep{li2022ringexchange}. Second, in the presence of significant magnetoelastic coupling in CoTiO$_3$ \citep{dey2021magnetostriction, Hoffmann2021highfield}, extended domain walls are preferred over sharp ones as the latter require a discrete change in lattice parameter and hence a large elastic energy cost.

Interestingly, a recent Raman scattering measurement observed sharp ZC magnon peak at zero field\citep{li2022ringexchange}, which showed no change across the spin reorientation transition. Given the large difference in the probed volume by neutrons and photons at optical wavelengths (the penetration depth of the wavelengths used in Ref.~\citep{li2022ringexchange} is estimated to be only about 14~nm using the complex refractive index calculated in Ref.~\citep{ZHANG20156}), the magnon peak measured by Raman scattering is likely associated with only one magnetic domain. The results in Ref.~\citep{li2022ringexchange} therefore provide indirect evidence for the proposed explanation involving domain walls, and strongly rule out any intrinsic damping of magnons at the zone center. A direct evidence for our explanation could be found through position sensitive imaging and spectroscopy techniques to correlate the local spin directions with their low energy excitations.    

\subsection{Dirac Magnon}\label{Diracmagnondiscussion}
Within the bond-dependent model proposed in Ref.~\citep{Elliot2021},  the magnon spectra near the $\mathbf{K}$ point are strongly dependent on the ordered moment directions. Since the ordered moment directions are clearly altered across the spin reorientation transition, we argued in Section~\ref{sec2B} that it is hard to reconcile the observation of very similar magnon spectra at 0~T and 8~T with large bond-dependent interactions in CoTiO$_3$. As we show in the Supplemental Materials, although we could not completely rule out the bond-dependent interactions as the reasons for the observed $\mathbf{K}$-point gap, in order to explain the observed magnon spectra at \textit{both}  0~T and 8~T, the bond-dependent model requires certain \textit{ad hoc} assumptions that are hard to be justified physically.

Given the obvious caveats of the bond-dependent model, we now consider an alternative scenario where a magnon gap at the $\mathbf{K}$ point is created by fully gapping out the Dirac cone. As shown in Ref.~\citep{Elliot2021, ArunKreinPRB2022}, the existence of Dirac nodal lines is protected by the magnetic inversion symmetry of the magnon Hamiltonian, which is a combination of spatial inversion across an inversion center between the two neighbouring Co$^{2+}$ ions along the $\mathbf{c}$-axis, followed by time reversal that flips the ordered moment direction. This symmetry has to be broken to gap out the Dirac cone. This could happen either by a canting between the two neighbouring honeycomb planes, or a structural distortion of the crystal that makes the two octahedra directly on top of each other inequivalent. Since the canting, if existed, is clearly modified across the spin reorientation transition in CoTiO$_3$, observation of almost identical spectra at 0~T and 8~T rules out the first mechanism, and suggests the second mechanism to be the most plausible explanation for the absence of magnetic inversion symmetry and consequently a gap at the $\mathbf{K}$ point. 

In fact, indirect evidence for a crystal symmetry lower than the reported $R\bar{3}$ space group is found in a recent angle-dependent magnetic susceptibility measurement at room temperature, well into the paramagnetic regime \citep{Hoffmann2021highfield}. In their work, the authors found the susceptibility as a function of in-plane field direction to exhibit two-fold periodicity that is inconsistent with the $\mathcal{C}_3$ symmetry of the reported space group, which should show an isotropic in-plane susceptibility. Although a structural distortion has not been directly observed in X-ray or neutron powder diffraction measurements, it cannot be ruled out considering the low sensitivity of powder diffraction measurement (especially in the case of X-ray powder diffraction) to weak symmetry forbidden reflections with small changes in the structure of the oxygen octahedra. A future single crystal synchrotron or neutron diffraction experiment is desired to directly observe such a distortion. If the observed Dirac gap is indeed due a small structural distortion in CoTiO$_3$, its size should be quite sensitive to a change in the crystal structure. An inelastic neutron scattering experiment on a crystal under an in-plane strain, combined with a single crystal diffraction measurement, can therefore provide direct evidence for the proposed mechanism.

\section{Conclusions}
In conclusion, we have carried out field-dependent high-resolution inelastic neutron scattering on CoTiO$_3$ to study the changes of its zone center and Dirac magnon across a spin reorientation transition induced by an in-plane magnetic field. Through elastic neutron scattering measurements, we confirmed that our sample transitions from a multi-domain state at zero field to a high-field mono-domain state with a canted magnetic structure. Concurrent with the spin orientation transition, we found a large change in the energy and the width of the zone center magnon peak, the latter of which was argued to be consistent with an unusual zero-field state with extended domain walls. The complex domain configuration at 0~T uncovered in the present work highlights the difficulty in the determination of the exchange parameter based solely on the zero-field data, and the importance of our field-dependent measurements. In contrast with the behaviours of the zone center magnon, we found the magnon spectra near the $\mathbf{K}$ point to be almost unaffected by a change in ordered moment directions and the domain structure. This observation is difficult to explain within the framework of the bond-dependent model, which predicts a strong dependence of the magnon spectra on the ordered moment directions. We argued that a symmetry-lowering structural distortion appears to be a more likely explanation for the observed gap. Although this structural distortion is yet to be directly observed in a diffraction experiment, we believe our work will provide strong constraints for any models proposed for CoTiO$_3$ in the future.

\section{Acknowledgements}
Work at the University of Toronto was supported by the Natural Science and Engineering Research Council (NSERC) of Canada through Discovery Grant No. RGPIN-2019-06449, the Canada Foundation for Innovation, and the Ontario Research Fund. G.J.S. acknowledges the support provided by MOST-Taiwan under Project No. 105-2112-M-027-003-MY3. We acknowledge the support of the National Institute of Standards and Technology, U.S. Department of Commerce, in providing the neutron research facilities used in this work.  This research used resources at the Spallation Neutron Source, a DOE Office of Science User Facility operated by the Oak Ridge National Laboratory. Use of the MAD beamline at the McMaster Nuclear Reactor is supported by McMaster University and the Canada Foundation for Innovation.

\FloatBarrier
\end{document}